\documentclass[journal]{IEEEtran}
\ifCLASSINFOpdf
\usepackage{graphicx}
\usepackage{epstopdf}
\usepackage{amsthm}
\usepackage{amsmath} 

\usepackage{amssymb}

\newtheorem{definition}{Definition} 

\newtheorem{prop}{Proposition}   

\DeclareMathAlphabet{\mathpzc}{OT1}{pzc}{m}{it}
\DeclareFontFamily{OT1}{pzc}{}
\DeclareFontShape{OT1}{pzc}{m}{it}{<-> s * [0.900] pzcmi7t}{}
\DeclareMathAlphabet{\mathpzc}{OT1}{pzc}{m}{it}

\usepackage{xcolor} 

\usepackage{color, colortbl}
\definecolor{Gray}{gray}{0.9}
\definecolor{LightCyan}{rgb}{0.88,1,1}

\usepackage{threeparttable}

\usepackage{caption}

\usepackage[noadjust]{cite}

\usepackage{subcaption}    

\usepackage{float} 

\usepackage{array}
\newcolumntype{L}[1]{>{\raggedright\let\newline\\\arraybackslash\hspace{0pt}}m{#1}}
\newcolumntype{C}[1]{>{\centering\let\newline\\\arraybackslash\hspace{0pt}}m{#1}}
\newcolumntype{R}[1]{>{\raggedleft\let\newline\\\arraybackslash\hspace{0pt}}m{#1}}

\usepackage{algorithm} 
\usepackage{algpseudocode}
\def\NoNumber#1{{\def\alglinenumber##1{}\State #1}\addtocounter{ALG@line}{-1}}

\usepackage{pgfplots}
\usepackage{lipsum,adjustbox}    
\usetikzlibrary{patterns}        

\pgfplotsset{compat=1.12}       

\else
\fi
\begin{document}
%
\title{Decentralized and stable matching in Peer-to-Peer energy trading\vspace{-3mm}}
%
%
%

\vspace{-2mm}\author{Nitin ~Singha,~\IEEEmembership{Member,~IEEE},~V~Shreyas,~Sandeep ~Kumar
\thanks{N. Singha is with the Department
of Electronics and Communication Engineering, NIT Delhi,
India, e-mail: nitinsingha@gmail.com\\
V Shreyas and Sandeep Kumar are with Department
of Electrical Engineering, IIT Delhi,
India }
}

%
%

\markboth{Journal of \LaTeX\ Class Files,~Vol.~14, No.~8, August~2015}%
{Shell \MakeLowercase{\textit{et al.}}: Bare Demo of IEEEtran.cls for IEEE Journals}
%



\maketitle

\begin{abstract}
In peer-to-peer ({\rm P2P}) energy trading, a secured infrastructure is required to manage trade and record monetary transactions. A central server/authority can be used for this. But there is a risk of central authority influencing the energy price. So blockchain technology is being preferred as a secured infrastructure in {\rm P2P} trading. Blockchain provides a distributed repository along with smart contracts for trade management. This reduces the influence of central authority in trading. However, these blockchain-based systems still rely on a central authority to pair/match sellers with consumers for trading energy. The central authority can interfere with the matching process to profit a selected set of users. Further,  a centralized authority also charges for its services, thereby increasing the cost of energy. We propose two distributed mechanisms to match sellers with consumers. The first mechanism doesn't allow for price negotiations between sellers and consumers, whereas the second does. We also calculate the time complexity and the stability of the matching process for both mechanisms. Using simulation, we compare the influence of centralized control and energy prices between the proposed and the existing mechanisms. The overall work strives to promote the free market and reduce energy prices.
\end{abstract}

\begin{IEEEkeywords}Distributed system, electricity market, game theory, peer-to-peer energy trading, stable matching.
\end{IEEEkeywords}

%
\IEEEpeerreviewmaketitle

\section{Introduction}
\vspace{-2mm}
To meet society's burgeoning demand for energy, its local production is the need of hour \cite{renewable_energy,local_trading,decentralize_market,intro_P2Ptrading}. A trading platform is required to facilitate the local exchange of energy. A Peer-to-peer ({\rm P2P}) energy trading network is an online marketplace to trade energy between users within a small geographical area \cite{intro_P2Ptrading1,intro_P2Ptrading2,survey_smart_grid,P2Ptrading_Elecbay}. For {\rm P2P} trading, a secure platform is required to record monetary transactions, handle price negotiations and resolve conflicts during the trading. Using a single/central authority to perform these functions has several limitations, \textit{viz.}, a single point of failure, privacy issues like confidential data leaks. Further, a centralized authority can also influence energy prices by modifying trading records. Thus, blockchain technology is being preferred over the centralized model in P2P energy trading \cite{NERA_review}. Blockchain stores data in a distributed fashion in such a way that data is virtually immutable and can be accessed by authorized personnel only. It also provides smart contracts to handle trading issues. 

\textbf{Related Work.} Work in \cite{dec_smartgrid} presents a blockchain model that enable users to perform anonymous and secure transactions during energy trading. The blockchain model in \cite{Energy_Coin} uses consortium blockchain and replaces centralized authority with a set of supernodes (or auctioneers) to facilitate trading between users. The blockchain has also been integrated with other services to improve security and increase the frequency of trading. Work in \cite{cloud_energy} combines blockchain with a cloud-based energy management service to create a more flexible, secure, and low-cost system. Work in \cite{Energy_loan} extends loan service to users of blockchain to increase the number of transactions in the network. 

Although blockchain-based systems have decentralized many functions, but they still require a central authority to group sellers with consumers\cite{dec_smartgrid,latest_P2P,Energy_Coin}. The centralized control can interfere with group formation to profit a specific set of consumers or sellers. Blockchain model proposed in \cite{Energy_Coin}, uses an auctioneer that groups sellers with consumers based on supply and demand of energy.  Models proposed in \cite{dec_smartgrid} and \cite{latest_P2P} use a public repository to store details about energy offered by sellers and its price.  Using these details, consumers approach sellers for energy. While trading model in \cite{dec_smartgrid} does not allow for negotiation of prices, model in \cite{latest_P2P} does. The auctioneer in \cite{Energy_Coin}, and hosting entities (electricity utility company or grid operator) in \cite{dec_smartgrid} and \cite{latest_P2P} can manipulate trading data to influence energy prices. Price manipulation is against the free-market principle, where only supply and demand (without external influence) should decide energy price.

To minimize centralized control, the auctioneer has been automated using a smart contract in the blockchain model \cite{autction_smart_grid}. But such modeling has been limited to a simple auction mechanisms like sealed bid auction. To the best of our knowledge, the complex auctioning mechanisms used in P2P energy trading have never been automated. Further auction automation reduces the speed of the system. Another alternative to centralized control can be the use of distributed/decentralized algorithms like the alternating direction method of multipliers ({\rm ADMM}) \cite{ADMM}. {\rm ADMM} is widely used in energy hubs inside smart buildings to solve load balancing problems. However, there is no decentralized algorithm to group sellers with consumers in P2P trading.
	
	\textbf{Motivation.} To reduce external interference on price, we require a decentralized algorithm for grouping sellers with consumers. However, decentralized group formation can create unstable grouping \cite{college_admission}. There will exist pairs of consumers and sellers who prefer each other over their current matching. They will break their current pairing to form pair with the preferred one. Due to new pair formation, some of the already matched users become unmatched, and again group formation procedure has to be restarted. To avoid this, many existing decentralized mechanisms exist that create stable groups \cite{college_admission,hospital_matching,goods_matching}. These matching mechanisms generate primarily two types of group formation: i) one-one matching: a single member of one group is matched to a single member of another group \cite{college_admission} and ii) one-many matching: a single member of one group is matched to many members of another group \cite{college_admission,hospital_matching,goods_matching}. Many modified grouping mechanisms that also consider the minimum and maximum quota for the individuals also exist \cite{goods_matching,min_quota,other3}. 

Although these mechanisms generate stable group formation, they cannot be directly applied to {\rm P2P} energy trading. In a {\rm P2P} scenario, a seller can sell its energy to multiple consumers, and consumers can purchase energy from multiple sellers simultaneously, \textit{i.e.}, many-to-many matching. Further, the amount of energy available with different sellers is different, which changes the number of consumers that can be matched with each seller. Similarly, the number of sellers matched with each consumer is different due to different energy requirements. Hence, we require a new decentralized algorithm to group sellers with consumers in the P2P energy trading network. 

\textbf{Contribution.} The main contributions are as follows.\\
1)  A framework is presented for decentralized group formation between sellers and consumers in {\rm P2P} trading network.\\
2) We propose two decentralized  algorithms, \textit{viz.}, energy matching ({\rm EM}) and negotiable energy matching {\rm NEM} algorithm for grouping sellers with consumers. While {\rm NEM} allows price negotiation, {\rm EM} does not. These algorithms operate in the decentralized fashion, \textit{i.e.} they are executed independently at every member in the network.\\ 
3) We also provide convergence proof and establish that the group formation is stable for the above algorithms.\\
4) The effectiveness of the proposed algorithms is extensively verified through a variety of experiments. The results validate that both energy price and the impact of centralized control on price are reduced by the proposed algorithms with respect to existing state-of-the-art algorithms \cite{dec_smartgrid,Energy_Coin}.

Hence, {\rm EM} and {\rm NEM} algorithms promote free-trade. This helps in creating a fair marketplace for {\rm P2P} energy trading. 
   
 \textbf{Notation.} A symbol in calligraphic font represents a set (e.g., $\mathcal{S}_{a}$ implies a set of sellers). A scalar is represented by a small letter (e.g., x) and
 a vector is represented by bold-faced letter with “bar” (e.g.
 $\mathbf{\bar{x}}$). A bold faced letter in upper case (e.g., $\mathbf{X}$) represents a matrix and  $x_{ij}$ represents the $(i, j)^{th}$ entry of matrix $\mathbf{X}$. The dimension of a matrix is omitted whenever it is clear from the context.   
 \vspace{-3mm}
\section{System Model and Problem Formulation}
\label{sec:model}
 \vspace{-2mm}
 This section presents a model for decentralized group formation in P2P trading. During trading, consumers buy energy blocks from the sellers. A block is the minimum amount of energy that can be traded. The size of the block can be varied in the system, depending on the minimum amount of energy members desire to trade. The trading is based on the preferences of sellers and consumers. This model includes the individual preferences of users. Further, this model simplifies matching process analysis by introducing the concept of virtual sellers and consumers. Virtual users transform many-to-many matching to one-to-one matching. Based on this model, we formulate the problem of energy matching. Energy matching is the process of grouping sellers with consumers for trading energy. We now introduce the model.
  \vspace{-3mm}
\subsection{Model}
 \vspace{-2mm}
\label{sec:Model}
\hspace{-4mm} P2P trading market can be represented by $(\mathcal{S}_{a}, \mathcal{C}_{a}, \mathcal{D}, \mathcal{B}, \mathcal{S}_{v}, \mathcal{C}_{v}, \mathcal{>_{S}}, \mathcal{>_{C}})$. Its components are described as follows.\\ \textbf{Sellers and consumers.} The sets $\mathcal{S}_{a}=\{s_1,\cdots,s_N\}$ and $\mathcal{C}_{a}=\{c_1,\cdots,c_M\}$ represents all the sellers and consumers, respectively, in the network. \\ 
\textbf{Energy availability and demand.} $\mathcal{D}=\{d_{s_1},\cdots,d_{s_N}\}$ and  $\mathcal{B}=\{b_{c_1},\cdots,b_{c_M}\}$ represents number of blocks available and demanded, respectively, in the network.  $d_{s_i}$ and $b_{c_i}$ are blocks associated with seller $s_i$ and consumer $c_i$, respectively.\\ 
\textbf{Virtual sellers and consumers.} Total number of energy blocks available and demanded in the P2P network are $S=\sum_{s_i{\in}\mathcal{S}_{a}}{d_{s_i}}$ and $C=\sum_{c_i{\in}\mathcal{C}_{a}}{b_{c_i}}$, respectively. We assume that each block available constitute a virtual seller and each block required constitute a virtual consumer. Sets $\mathcal{S}_{v}=\{1,2,\cdots,S\}$ and $\mathcal{C}_{v}=\{1,2,\cdots,C\}$ denote total number of virtual sellers and consumers in the network. To differentiate $\mathcal{S}_{a}$ and $\mathcal{C}_{a}$ from $\mathcal{S}_{v}$ and $\mathcal{C}_{v}$, $\mathcal{S}_{a}$ and $\mathcal{C}_{a}$ will be referred as actual sellers and actual consumers.\\
\textbf{Preference list.} The symbols $\mathcal{>_{S}}=\{>_{1},\cdots,>_{S}\}$ and $\mathcal{>_{C}}=\{>_{1},\cdots,>_{C}\}$ represent the preference list of all sellers and consumers, respectively. 
If a seller $i$ prefers consumer $j$ for $k$, then this relationship is represented as $j>_{i}k$. The preference list has the following properties.\\ 
$\bullet$ Preference relation is transitive. If seller $i$ prefers consumer $l$ over $m$, \textit{i.e.}, $l\mathcal{>}_{i}m$, and  it prefers consumer $m$ over $n$, \textit{i.e.}, $m\mathcal{>}_{i}n$. This implies that seller $i$ will prefer consumer $l$ over $n$, \textit{i.e.}, $l\mathcal{>}_{i}n$.\\
$\bullet$ Any seller $i$ has a strict preference relation $\mathcal{>}_{i}$ over set of consumers $\mathcal{C}_{v}$, and any consumer $j$ also has strict preference relation $\mathcal{>}_{j}$ over set of sellers $\mathcal{S}_{v}$. It implies that for a seller (or consumer) their can not be two or more consumers (or sellers) which are equally preferred by it.

The system proposed in this paper is generic and can work with any preference list. But for ease of understanding, we consider the following preferences. A consumer prefers sellers in increasing order of their selling price (seller with lowest selling price is given highest priority). In case of tie nearest seller is preferred to save on transmission cost\footnote{Greater the distance, higher is the infrastructure usage, thereby increasing transmission cost.} and losses\footnote{Energy losses increase with the distance of transmission.}. A seller prioritize consumers in the decreasing order of their bids (consumer with the highest bid is given the highest priority). In the case of a tie, the nearest consumer is preferred. \\
\textbf{Other components of model}: ${\forall}c_i{\in}\mathcal{C}_{a}$ and ${\forall}s_i{\in}\mathcal{S}_{a}$,  $\mathcal{A}_{c_i}$ and $\mathcal{A}_{s_i}$ are sets representing allocated and allocation lists, respectively. $\mathcal{A}_{c_i}=\{a^{c_i}_{s_1},\cdots,a^{c_i}_{s_N}\}$, where $a^{c_i}_{s_i}$ are the number of energy blocks allocated to consumer $c_i$ from seller $s_i$. $\mathcal{A}_{s_i}=\{a^{s_i}_{c_1},\cdots,a^{s_i}_{c_N}\}$, where $a^{s_i}_{c_i}$ is the number of energy blocks allocated by seller $s_i$ to consumer $c_i$.

$L_{c_i}$ denotes the latest seller who has promised energy to consumer $c_i$. $e_{c_i}^{mtc}$ and $e_{c_i}^{umtc}$ are number of energy blocks matched and unmatched with consumer $c_i$. $P_{c_i}$ is the unprocessed seller list that contains the list of sellers who have not been contacted for energy. $R_{s_i}$ is the list of current requesters of seller $s_i$. Current requesters are those consumers who have requested seller $s_i$ for energy and are waiting for its response. 
 \vspace{-3mm}
\subsection{Problem Formulation}
 \vspace{-1mm}
\hspace{-3.6mm}Energy matching implies grouping sellers with the consumers. This matching along with its attributes is defined as follows. \vspace{-2mm}
\begin{definition}[\textbf{Energy Matching}]For a given set of virtual sellers $\mathcal{S}_{v}$ and consumers $\mathcal{C}_{v}$, energy matching is a mapping ${\mu}:{\mathcal{S}_{v}}\cup{\mathcal{C}_{v}}{\rightarrow}{\mathcal{S}_{v}}\cup{\mathcal{C}_{v}}$, such that:
	\begin{enumerate}
		\item $\mu(i){\subseteq}\mathcal{C}_{v},~{\forall}i{\in}{\mathcal{S}_{v}}$: A seller can sell energy only to a consumer.
		\item $\mu(j){\subseteq}\mathcal{S}_{v},~{\forall}j{\in}{\mathcal{C}_{v}}$: A consumer can purchase energy only from a seller.
		\item For any seller $i$ consumer $j$, $j{\in}\mu(i){\iff}i{\in}\mu(j)$.
	\end{enumerate}
\textit{Note:} In terms of actual sellers $\mathcal{S}_{a}$ and consumers $\mathcal{C}_{a}$, the matching is defined as follows:
\begin{itemize}
	\item $\mu(s_i), {s_i}{\in}\mathcal{S}_{a}$, contains the name of consumers to which blocks of $s_i$ are sold. If $s_i$ sold $2$ blocks to $c_j$, and one block each to $c_{x}$ and $c_{y}$, then  $\mu(s_i)=\{c_j,c_j,c_x,c_y\}$.
	\item Similarly $\mu(c_i), {c_i}{\in}\mathcal{C}_{a}$ contains name of sellers from which blocks are purchased by $c_i$. 
\end{itemize} \vspace{-2mm}
\end{definition}
\begin{definition}[\textbf{Feasible Energy Matching}] An energy matching is feasible if number of blocks sold by the seller is not more than the blocks available with it, and the number of blocks awarded to consumer is not more than its requirement, \textit{i.e.}, $|\mu(s_i)|{\leq}d_{s_i},{\forall}s_i{\in}{\mathcal{S}_{a}}$, and $|\mu(c_i)|{\leq}b_{c_i},{\forall}c_i{\in}{\mathcal{C}_{a}}$.
\end{definition}	 \vspace{-2mm}
\begin{definition}[\textbf{Stable Energy Matching}] An energy matching is stable if there are no blocking pairs in the matching. 	
\end{definition} \vspace{-2mm}
\begin{definition}[\textbf{Blocking Pair}] The tuple (i,j), where $i$ is a seller and $j$ is a consumer is called blocking pair if they both prefer each other to their current choices, \textit{i.e.}, (i,j) is blocking pair if $i{\in}\mathcal{S}_{v}$ and $j{\in}\mathcal{C}_{v}$, such that $i>_{j}\mu(j)$ and $j>_{i}\mu(i)$ simultaneously.
\end{definition} \vspace{-2mm}
Sellers and consumers are selfish and rational, and will always look for better choices. If there is a blocking pair (i,j) in matching, then both seller $i$ and consumer $j$ will break their current matching and group together. In this way they can form group with user higher in the preference list. Once seller $i$ and consumer $j$ form a group, their previous matches will become unmatched. Hence the matching which contains at least one single blocking pair is not stable. 

In the subsequent sections, we propose two algorithms that create a feasible and stable matching between consumers and sellers. While designing algorithms, we have considered actual sellers $\mathcal{S}_a$ and actual consumers $\mathcal{C}_a$ instead of virtual sellers $\mathcal{S}_v$ and consumers $\mathcal{C}_v$. Virtual sellers and consumers are used for understanding the matching process, but practical trading occurs between actual sellers and consumers.
\section{How to perform matching}
\label{sec:EM}
\begin{algorithm}
		\caption{Energy Matching (EM) Algorithm}
    \textbf{Input:} Bids of $N$ sellers and $M$ consumers, blocks \hspace*{\algorithmicindent}available  $\mathcal{D}$$=$$\{d_{s_1},\cdots,d_{s_N}\}$, and blocks demanded \hspace*{\algorithmicindent}$\mathcal{B}$$=$$\{b_{c_1},\cdots,b_{c_M}\}$.\\ 
       	\textbf{Output:} Energy mapping $\mu$.
	\begin{algorithmic}[1]	
  \State\textbf{Initialization:}
	${\forall}$$c_i{\in}\mathcal{C}_{a}$, preference list $\mathcal{>_{C}}$=$\{c_1,\cdots,c_M\}$, \hspace*{\algorithmicindent}latest service provider $L_{c_i}$$=$$\phi$, matched energy blocks \hspace*{\algorithmicindent}$e^{\rm{mtc}}_{c_i}$$=$$0$,  unmatched energy blocks $e^{\rm{umtc}}_{c_i}$$=$$b_{c_i}$, and \hspace*{\algorithmicindent}unprocessed  list $\mathcal{P}_{c_i}\!\!=~>_{c_i}$, where $>_{c_i}$${\in}$$\mathcal{>_{C}}$.
	${\forall}s_i{\in}\mathcal{S}_{a}$, \hspace*{\algorithmicindent}current requester list $\mathcal{R}_{s_i}=\phi$.
	${\forall}$$c_i{\in}\mathcal{C}_{a}$, $c_i$ sets every \hspace*{\algorithmicindent}element of the allocated list $\mathcal{A}_{c_i}$ to $0$. ${\forall}s_i{\in}\mathcal{S}_{a}$, $s_i$ sets \hspace*{\algorithmicindent}every element of the allocation list	$\mathcal{A}_{s_i}$ to 0.
		\While{	(${\exists}$ $c_i$ who is unmatched) and $(\mathcal{P}_{c_i}{\neq}\phi)$}{
		\ForAll{unmatched consumers $c_i$ with $\mathcal{P}_{c_i}{\neq}\phi$} {
			\State \hspace{-4mm}$s_i$ = most preferred seller in $\mathcal{P}_{c_i}$. 
			\State \hspace{-4mm}$c_i$ requests $s_i$ for $e^{\rm{umtc}}_{c_i}$ blocks.
			\State \hspace{-4mm}Add $c_i$ to $s_i$'s current requester list, $\mathcal{R}_{s_i}$$=$$\mathcal{R}_{s_i}$${\cup}$$\{c_i\}$.
			\State \hspace{-4mm}Remove $s_i$ from $c_i$'s  unprocessed list, 
 $\mathcal{P}_{c_i}$$=$$\mathcal{P}_{c_i}$${\setminus}$$\{s_i\}$.
		\EndFor}
		\ForAll{sellers $s_i$ with $\mathcal{R}_{s_i}{\neq}\phi$}{
		\State \hspace{-4mm}$s_i$ updates its preference list.
		\State \hspace{-4mm}$s_i$ accepts up to $d_{s_i}$ requests from most preferred 
		 \NoNumber{\hspace{-4mm}requesters in $\mathcal{R}_{s_i}{\cup}\mu{(s_i)}$.} 
		\State \hspace{-4mm}$s_i$ updates its mapping list $\mu{(s_i)}$.
		\State \hspace{-4mm}$s_i$ updates its allocation list $\mathcal{A}_{s_i}$ 
		\State \hspace{-4mm}$s_i$ clears its current requester list, $\mathcal{R}_{s_i}=\phi$. 
		\EndFor	}
		\State Based on updated mapping list of sellers, all    \NoNumber{consumers update their mapping lists}
	    
		\State Based on updated mapping list of consumers,   \NoNumber{$L_{c_i}$, $\mathcal{A}_{c_i}$, $e^{\rm{mtc}}_{c_i}$, and $e^{\rm{umtc}}_{c_i}$ are updated {$\forall$}$c_i{\in}\mathcal{C}_{a}$} 
		\If{ Seller removed from updated consumer list  $~~~~~~~~~~~~~$ $\mu(c_i)$ is not latest service provider $L_{c_i}$}
		\State \hspace{-3mm}Update $c_i$'s unprocessed list $\mathcal{P}_{c_i}$$=$$\mathcal{P}_{c_i}$${\cup}$$\{L_{c_i}\}$.
		\EndIf
		\EndWhile
}
		\State\Return $\mu$
	\end{algorithmic}
	\label{alg:EM}
\end{algorithm}

\noindent We now propose an energy matching {\sf(EM)} algorithm. 
 The energy matching starts with sellers propagating information across the network that they want to sell energy and its selling price\footnote{Usually, sellers are reluctant to disclose their trading information, so they may not reveal the amount of energy available.}. A seller informs distributed system operator (\rm DSO)\footnote{{\rm DSO} is entity responsible for managing infrastructure used for energy transfer. \vspace{-2mm}}, its neighbors, and frequent consumers about its selling price. Consumers download a list of sellers from {\rm DSO}, neighbors, and previous sellers; consolidate this list to construct a preference list of sellers $\mathcal{>_{C}}$. Initially, only consumers know about their preference list $\mathcal{>_{C}}$, and the sellers do not know about their preference list. In {\rm EM} algorithm, a seller $s_j$ does not explicitly maintain its preference list; instead, it maintains a mapping list $\mu(s_j)$, containing the list of prospective consumers to which energy will be supplied. A seller $s_j$ dynamically updates its mapping list $\mu(s_j)$ based on the bids received by the consumers. Consumers start bidding for energy to sellers at the beginning of every round. The round\footnote{ The clock of sellers and consumers are synchronized with {\rm DSO} to start and end the round.\vspace{-2mm}} is a time period during which consumers submit their bid to a single seller, and sellers update their mapping list. 

In the first round of {\rm EM} algorithm (Algorithm \ref{alg:EM}), each consumer ($c_i{\in}\mathcal{C}_a$) will submit a bid to the seller ($s_j{\in}\mathcal{S}_a$) with the highest priority. After receiving bids from the consumers, a seller $s_j$ will store a list of prospective consumers in its mapping list $\mu({s_j})$. Seller $s_j$ will also update its allocation list $A_{s_j}$, specifying the number of blocks allocated to different requesters. The number of blocks allocated by seller $s_j$ cannot be greater than the blocks $d_{s_j}$ available with it. In case the requests received by seller $s_j$ are greater than $d_{s_j}$, $d_{s_j}$ blocks are allocated among the requesting consumers in decreasing order of their priority. After $d_{s_j}$ blocks are mapped, the remaining requests are rejected by $s_j$. Based on sellers' preferences,  consumers determine whether they will receive blocks from the requested seller at the end of the round. They also know how many blocks they will be receiving from the seller.

If all the energy blocks required by a consumer $c_i$ are matched/paired to a seller, then consumer $c_i$ will not bid in the second round; else, consumer $c_i$ will submit the bid to the seller that is second in its preference list. The number of blocks demanded from the second seller $s_k$ is equal to unmatched/unpaired blocks. If the seller $s_k$ has unmatched blocks equal to blocks requested, it will allocate them to consumer $c_i$. Otherwise, it will compare the bid of consumer $c_i$ with the existing consumers that have been promised blocks. If the consumer $c_i$ bid is higher than existing ones, then the seller $s_k$ will reallocate blocks to the consumer $c_i$. If a consumer has unmatched blocks after the second round, it will bid again in the subsequent round till all blocks are matched. 

Once all the blocks with consumer $c_i$ are matched, it may not require bidding in the current round, but it may have to bid in a later round if the seller that promised block receives a higher bid (buying price) than consumer $c_i$ in the future. Such unmatched consumer $c_i$ may not directly demand from the next unproposed seller in the preference list. If consumer $c_i$ becomes unmatched due to reallocation of energy block by seller other than the latest seller $L_{c_i}$ that promised the block, there are chances that the latest seller $L_{c_i}$ can provide energy. In the previous bid to seller $L_{c_i}$, consumer $c_i$ requested for lesser blocks, as recently unmatched blocks were matched at that time. So, if consumer $c_i$ becomes unmatched due to seller other than latest seller $L_{c_i}$, it will first demand from the seller $L_{c_i}$. If consumer $c_i$ is still unmatched, it will start requesting unproposed sellers in its preference list.

In this way, both consumers and sellers continue to update their mapping list (prospective trading partners) in every round. This algorithm runs until all the consumers have been matched or the unmatched consumers have exhausted their preference list. 
	\subsection{Toy Example of Energy Matching}
	\begin{figure*}
			\begin{subfigure}{\textwidth}
			\centering\includegraphics[width=125mm,height=60mm, keepaspectratio=false]{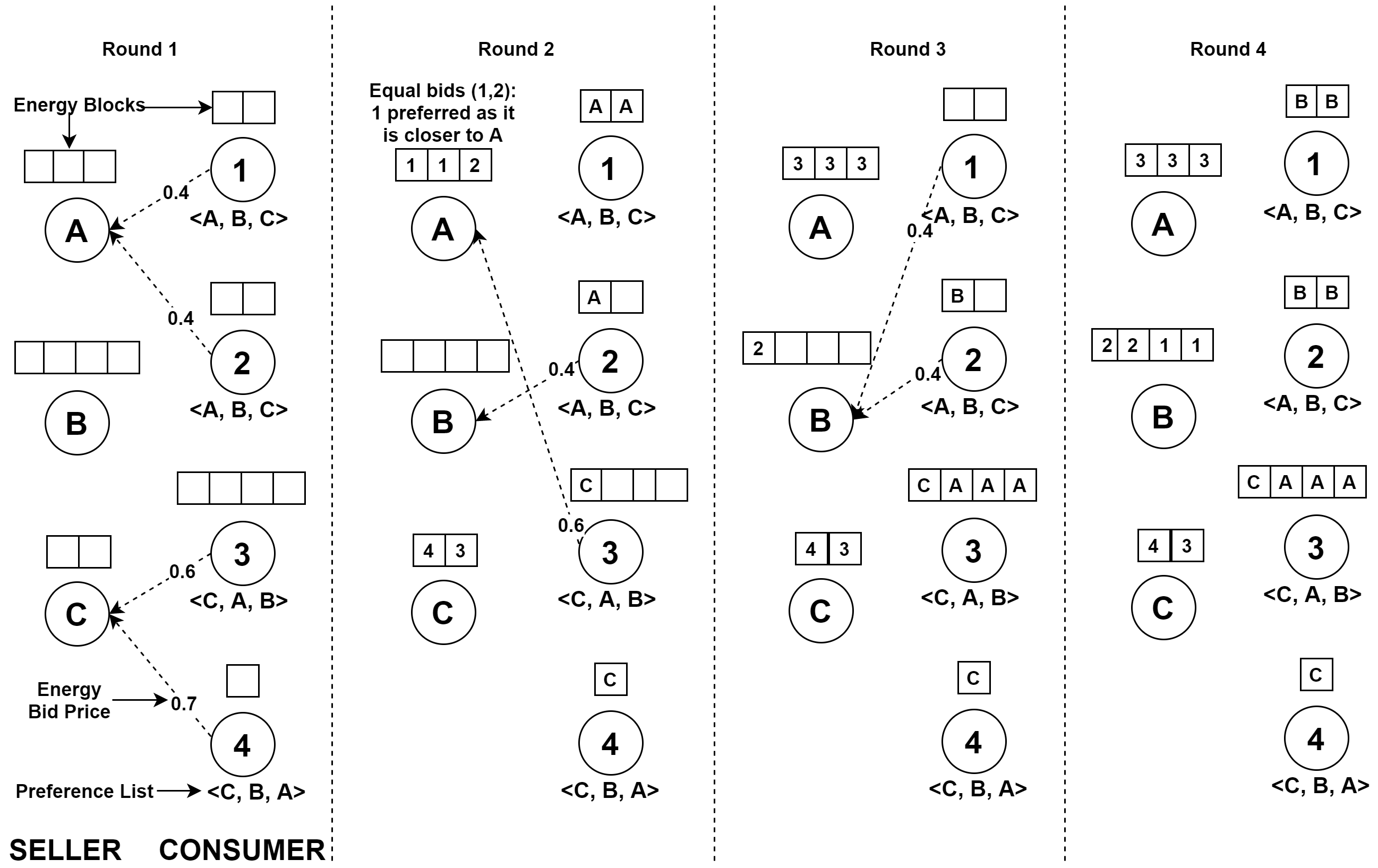}
			\vspace{-3mm}\caption{Blocks to be sold = Blocks demanded.}
			\label{fig:symm_matching}
		\end{subfigure}
		\newline
		\begin{subfigure}{\textwidth}
			\centering\includegraphics[width=125mm,height=60mm, keepaspectratio=false]{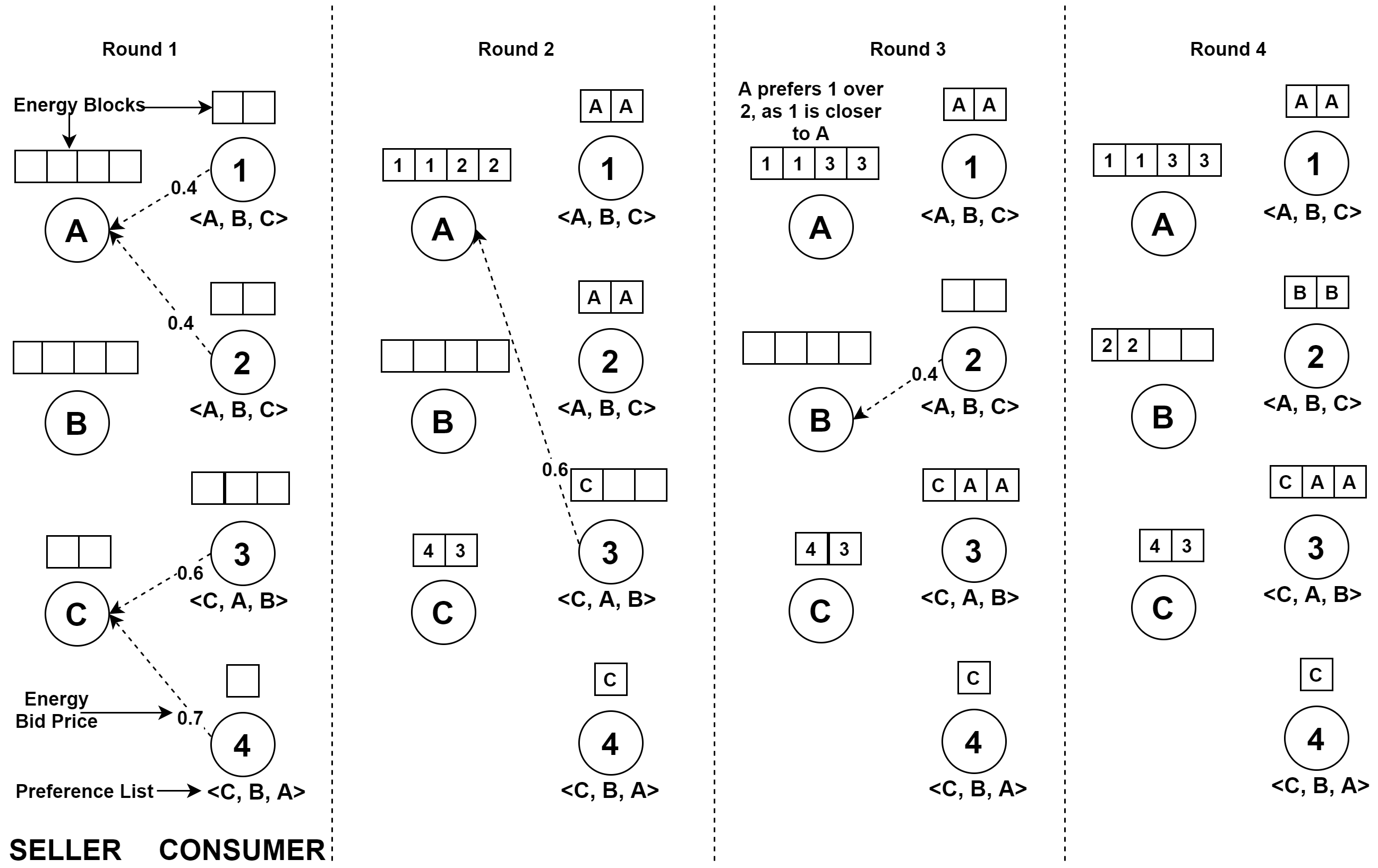}
			\vspace{-3mm}\caption{Blocks to be sold $>$ Blocks Demanded.}
			\label{fig:seller_matching}
		\end{subfigure}
		\newline
		\begin{subfigure}{\textwidth}
			\centering\includegraphics[width=125mm,height=60mm, keepaspectratio=false]{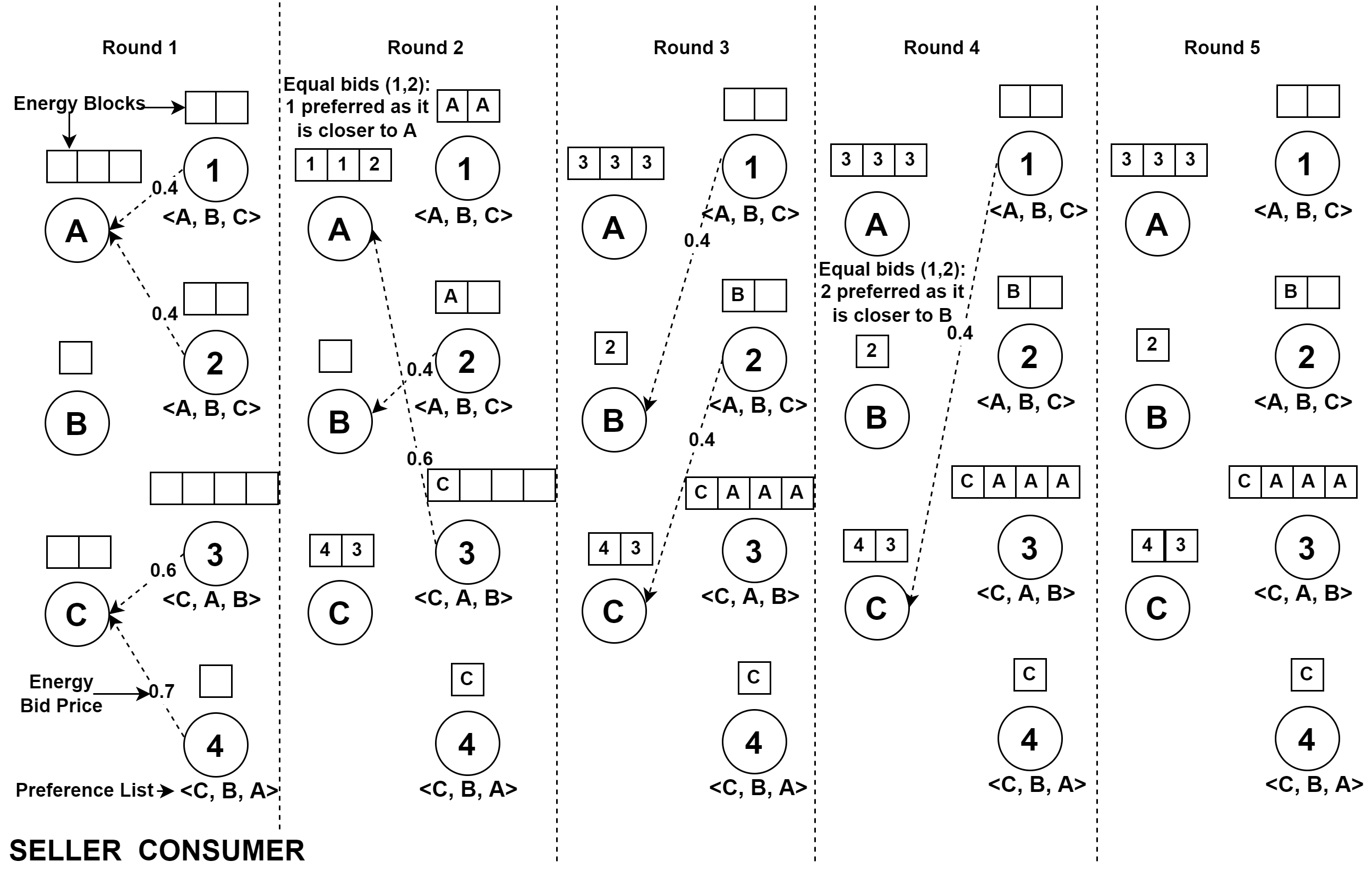}
			\vspace{-5mm}\caption{\centering Blocks to be sold $<$ Blocks demanded.}
			\label{fig:consumer_matching}
		\end{subfigure}
	\vspace{-2mm}
		\caption{\centering Energy Matching: Matching sellers with consumers.}  		
		\label{fig:matching}
		\vspace{-7mm}
	\end{figure*}
	As shown in Fig. \ref{fig:matching}, there are $4$ consumers $1$, $2$, $3$, and $4$; and $3$ sellers $A$, $B$, and $C$. The energy demand of consumers and supply by sellers is represented in terms of blocks, e.g., in Fig. \ref{fig:symm_matching}, seller $A$ is selling $3$ blocks of energy, whereas consumer $1$ requires two blocks of energy. 

The {\rm EM} algorithm matches energy blocks of sellers with  consumers. These energy blocks are matched based on the preferences of their respective consumers and sellers.\\ 
\textbf{Consumer's preference.} For simplicity, we assume that all sellers in Fig. \ref{fig:matching} are selling at the same price. Seller A is nearest, and seller C is farthest from consumer $1$. Consumer $1$ will give the highest preference to seller $A$, then $B$, and least priority to seller $C$. The preference list of consumer $1$ is represented as $<A,B,C>$ in Fig. \ref{fig:matching}. The preference list of consumer $2$, $3$, and $4$ are $<A,B,C>$, $<C,A,B>$, and $<C,B,A>$, respectively.\\
\textbf{Seller's preference.} Let price bids of consumers for each block is given by $\mathit{\mathbf{B_{\rm con}}}=\{0.4,0.4,0.6,0.7\}$, \textit{i.e.}, bids of consumer $1,~2,~3,~\text{and}~4$ are $0.4,0.4,~0.6,~\text{and}~0.7$ dollars, respectively. Bids of consumer $1$ and $2$ are same. In such case, the seller prefers the nearest consumer. Consumer $1$ is nearer to seller A, while consumer $2$ is nearer to $B$ and $C$. 
		
Unlike consumers, the seller's preference list is not available with the seller at starting of the matching process, as they do not know the bids of consumers. These lists get updated dynamically as the bids arrive at the seller. After all the bids have arrived, the final preference list for sellers $A$, $B$, and $C$ are $<4,~3,~1,~2>$, $<4,~3,~2,~1>$, and $<4,~3,~2,~1>$, respectively .

The energy matching starts with sellers propagating information about energy, and then consumers start bidding for energy. Apart from bids, the supply and demand across the network influence the matching process. There can be three different supply and demand scenarios discussed as follows.\\
	\textit{1) Number of blocks to be sold is equal to blocks requested: }Fig. \ref{fig:symm_matching} shows energy matching when an equal number of blocks $(9)$ are available with both sellers and consumers. During the first round, a consumer submits a bid to the seller at the top of its preference list. Consumers  $1$ and $2$, request $2$ blocks each from seller $A$, whereas  consumer $3$ and $4$, requests  $4$ and $1$ block, respectively from  seller $C$. Seller $A$ has only $3$ blocks, but it receives request for $4$ blocks. As seller $A$ can't meet demands of all requesters, so it allocates block to requesters in decreasing order of its preference. Seller $A$ allocates $2$ blocks to consumer $A$, and $1$ block to consumer $B$.  Similarly  seller $C$ prefers consumer $4$ over consumer $3$ during the allocation. After the first round, a prospective consumer for a given seller's block is shown by the name of that consumer on that block (refer round $2$ in Fig. \ref{fig:symm_matching}). Similarly, for consumers, the prospective seller is represented by the seller's name on the block. As blocks of consumers $2$ and $3$ remain unmatched during the first round, they again bid for energy in the second round.  In this round, seller $A$ receives a request from consumer $3$, that is more preferred by it than current prospective consumers. So it reallocates its blocks to consumer $3$.  Since consumers $1$ and $2$ become unmatched, they submit their bid to the next preferred seller $B$ in their preference list. The blocks of all the consumers get matched in round $4$, and the matching algorithm terminates.\\
	\textit{2) Number of blocks to be sold is greater than requested:} Sellers have $10$, whereas consumers request $8$ blocks (Fig. \ref{fig:seller_matching}). Same process repeats as discussed above. The algorithm terminates when all the blocks of consumers get matched, but there will be some unmatched blocks with sellers.\\ 
	\textit{3) Number of blocks to be sold is less than requested:} Sellers have $6$, and consumers request $9$ blocks (Fig. \ref{fig:consumer_matching}). Initially, the same process repeats as above. After the second round,  blocks with consumers $3$ and $4$ get matched, whereas some blocks of consumers $1$ and $2$ remain unmatched. During subsequent rounds, unmatched consumers send bids to the remaining sellers in their preference list. The algorithm terminates after the fourth round because unmatched consumers have exhausted their options. The matching completes with some unmatched blocks with consumers $1$ and $2$. \vspace{-2mm}
\begin{prop}[\textbf{Computational Complexity}]	The energy matching algorithm converges with the computational complexity of $\mathcal{O}(NML_{\rm max}{\tau}_{\rm max})$, where $N$, $M$ are the number of actual consumers and sellers, respectively. $L_{\rm max}$ is the maximum number of blocks demanded by any consumer, and ${\tau}_{\rm max}$ is the maximum time taken by any seller to allocate blocks in the network. \vspace{-2mm} 
	\label{prop:1}
\end{prop}\vspace{-1mm}
	\noindent To calculate computational complexity, we evaluate the performance of {\rm EM} algorithm during the worst case. Every consumer will call every seller in the network, and each consumer will demand the maximum number of blocks $L_{\rm max}$. Initially consumer $c_i$ calls a seller $s_j$ at top of preference list in first round for $L_{\rm max}$ blocks, and $s_j$ will allocate $L_{\rm max}$ blocks to the consumer $c_i$. In the worst case, at the end of a each subsequent round, $s_j$ will reallocate one block promised to $c_i$ to some other consumer. Consumer $c_i$ may request the next seller $s_k$ in the preference list for each unmatched block. Thus, consumer $c_i$ will request seller $s_k$, $L_{\rm max}$ times. Similarly, like seller $s_j$, seller $s_k$ also starts reallocating one block per round to another consumer. Thus the third seller in the preference list also needs to be contacted $L_{\rm max}$ times, and this process continues till all the sellers $M$ in the preference list are exhausted. Hence, in the worse case, the maximum number of requests made by any consumer $c_j$ to all the sellers is $((M-1)L_{\rm max})+1$; after that, {\rm EM} algorithm will terminate. 
	
	To form a matching between all the sellers $M$ with consumers $N$, the maximum number of requests generated are $N(((M-1)L_{\rm max})+1))$. If ${\tau}_{\rm max}$ is the maximum time taken by any seller to allocate block to a consumer, then the maximum time taken by {\rm EM} algorithm to group all the consumers with sellers in the network is $N(((M-1)L_{\rm max})+1)){\tau}_{\rm max}$. Asymptotically, this value converges to $NML_{\rm max}{\tau}_{\rm max}$. Hence computational complexity of {\rm EM } algorithm is $\mathcal{O}(NML_{\rm max}{\tau}_{\rm max})$.   \vspace{-2mm}
	\begin{prop}[\textbf{Feasibility and Stability}] The matching result of the EM algorithm is both feasible and stable.
		\label{prop:2}
	\end{prop} \vspace{-2mm}
    \noindent\textbf{Feasibility.} The EM algorithm is feasible. It does not allow sellers to sell blocks more than available and does not allow consumers to request blocks more than their requirement.
    
    \noindent \textbf{Stability.} We prove the stability of {\rm EM } algorithm by contradiction. For ease of understanding, we consider virtual consumers $i{\in}\mathcal{C}_v$, and sellers $j{\in}\mathcal{S}_v$ (refer section \ref{sec:Model} for concept of virtual users). Let us assume that matching achieved by {\rm EM} algorithm is not stable. Thus, there exists at least one blocking pair $(i,j)$  of seller $i$ with consumer $j$ in the matching, \textit{i.e}, consumer $j$ prefers to purchase from the seller $i$ than its current match $\mu(j)$ ($i>_{j}\mu(j)$), and at same time seller $i$ also prefers to sell to consumer $j$ than its current match $\mu(i)$ ($j>_{i}\mu(i)$). However, since they have been grouped by {\rm EM} algorithm, and consumer $j$ prefers seller $i$ over its current matched seller $\mu(j)$, so the consumer $j$ must have contacted seller $i$ before seller $\mu(j)$ during the execution of the algorithm. At that time, seller $i$ would have rejected the bid because it would have been matched to some other consumer $k$ higher in the preference list than consumer $j$ ($k>_{i}j$).  In {\rm EM} algorithm, only a seller can reallocate its energy block; and it will only reallocate when it receives bids from the consumer higher in the preference list than the current consumer. Thus the current consumer $\mu(i)$ to which seller $i$ is matched, has preference greater than consumer $k$, \textit{i.e.}, $\mu(i)>_{i}k$. Also $k>_{i}j$; and since preferences are transitive, it implies that $\mu(i)>_{i}j$  This is in contraction to earlier assumption $j>_{i}\mu(i)$, \textit{i.e.}, seller $i$ prefers  consumer $j$ over its currently matched consumer $\mu(i)$. Therefore, our initial assumption that a blocking pair $(i,j)$ exists is wrong; and the matching obtained using {\rm{EM}} algorithm is stable.  Hence, {\rm{EM}} algorithm converges in real time and generates a feasible and stable mapping. 
    \vspace{-4mm}
\section{Energy trading with negotiable energy prices}
 \vspace{-1mm}
\label{sec:NEM}
\begin{algorithm}[t]
	\caption{Negotiable Energy Matching ({\rm NEM}) Algorithm.}
 \textbf{Input:} Bids of $N$ sellers and $M$ consumers, blocks \hspace*{\algorithmicindent}available  $\mathcal{D}$$=$$\{d_{s_1},\cdots,d_{s_N}\}$, and blocks demanded \hspace*{\algorithmicindent}$\mathcal{B}$$=$$\{b_{c_1},\cdots,b_{c_M}\}$.\\ 
	\textbf{Output:} Energy mapping $\mu_{final}$ and $price$
	\begin{algorithmic}[1]	
		\State\textbf{Initialization:} $\mathcal{S}_{crnt}=\mathcal{S}_{a}$, $\mathcal{C}_{crnt}=\mathcal{C}_{a}$. Using \eqref{eq:selling_incriment} and  \hspace*{\algorithmicindent}\eqref{eq:buying_incriment}, calculate $\delta_{s_j}$ change in selling price for seller \hspace*{\algorithmicindent}$s_j$, and $\delta_{c_i}$ change in buying price for consumer $c_i$ \hspace*{\algorithmicindent}after one iteration. $itr=0$, and an $M{\times}N{\times}{K}$ matrix,  \hspace*{\algorithmicindent}where  $K= \max{d_{s_i},{\forall}d_{s_i}{\in}\mathcal{D}}$. $price$ initialized to $0$.  	  
		\While{	$(itr{<}T)$ and $(\mathcal{S}_{crnt}{\neq}\phi)$ and $(\mathcal{C}_{crnt}{\neq}\phi)$ }{
			
			\State \hspace{-5mm}For consumers in $\mathcal{C}_{crnt}$, and sellers in $\mathcal{S}_{crnt}$ run {\rm EM} algorithm.
			\ForAll{ consumers $c_i$ $\in$ $\mathcal{C}_{crnt}$}{
				\ForAll{ sellers $s_j$ $\in$ $\mathcal{S}_{crnt}$}{
					\If{$c_i$ matched with $s_j$ }	
					\If{$c_i$'s buying price $\geq$ $s_j$'s selling price {\color{white}a~~~~~~~~~~~~~~~} }
					\State Energy sold at price equal to average \NoNumber{of bids by $s_i$ and $c_j$.}
					\State Update  $\mu_{final}$ and current trading price \NoNumber{in  $price(i,j,x)$. $x=1:k$, and $k$ is the} \NoNumber{number of blocks promised by $s_j$ to $c_i$.}
					\EndIf
					\EndIf
					\EndFor} 	
				\EndFor} 	
			\State Sellers and consumers in $\mathcal{S}_{crnt}$ and $\mathcal{C}_{crnt}$ who have \NoNumber{traded their energy are removed from these sets}.
			\State Update bids of sellers and consumers \NoNumber{${\forall}s_j{\in}\mathcal{S}_{crnt},$ New bid = Current bid - $\delta_{s_j}$}, \NoNumber{${\forall}c_i{\in}\mathcal{C}_{crnt},$ New bid = Current bid + $\delta_{c_i}$} 
			\State $itr=itr+1$.	
			\EndWhile
		}			
		\State\hspace{-5.1mm}\Return  $\mu_{final}$ and $price$	
	\end{algorithmic}
	\label{alg:negotiable_EM}
\end{algorithm}
{\rm{EM}} algorithm discussed earlier does not allow sellers and buyers to negotiate the price of energy. For price negotiation, we propose negotiable energy matching {\rm{NEM}} algorithm (Algorithm \ref{alg:negotiable_EM}). In {\rm{NEM}} algorithm, the {\rm{EM}} algorithm is run repeatedly for $T$ iterations\footnote{Before start of {\rm NEM} algorithm, the sellers and consumers decide the value of T.} with different bids from members. Round and iteration are different in this paper. During rounds, bids of consumers and sellers remain the same, while after each iteration, bids change. {\rm EM} algorithm runs for multiple rounds, whereas {\rm NEM} algorithm runs for multiple iterations. In each iteration of {\rm NEM} algorithm, multiple rounds of {\rm EM} algorithm are executed.

At the start of first iteration in {\rm NEM} algorithm, sellers propagating their selling prices across the network (Algorithm \ref{alg:negotiable_EM}). After receiving the selling prices, consumer $c_i$ creates its preference list in a similar fashion as in {\rm EM} algorithm (refer section \ref{sec:EM}). Then sellers and consumers in the network are added in current sellers $\mathcal{S}_{crnt}$ and consumers $\mathcal{C}_{crnt}$ list, respectively. The list, along with other parameters (refer Algorithm \ref{alg:negotiable_EM}) is passed on to the {\rm EM} algorithm to create a matching $\mu$. In this matching, only those sellers will sell energy, who receive bids (buying price) higher than their selling price. Such pair of sellers and consumers are updated in the final mapping list $\mu_{final}$, along with the energy price in the final price list $price$. At the end of the iteration, sellers and consumers, who have traded their entire energy in the current iteration are removed from current sellers $\mathcal{S}_{crnt}$ and consumers $\mathcal{C}_{crnt}$ lists. 

In subsequent iterations, the same process is followed with updated selling and buying prices. Sellers and consumers will change their selling and buying price to increase their probability of trading energy. After each iteration, a seller $s_j$ reduces its selling price by amount $\delta_{s_j}$, whereas a consumer $c_i$ increases its buying price by amount $\delta_{c_i}$. ${\rm NEM}$ algorithm allows sellers and consumers to decide their $\delta_{s_j}$ and $\delta_{c_j}$, respectively. But for simulation purposes we have also proposed a method to calculate $\delta_{s_j}$ and $\delta_{c_j}$ in section \ref{subsec:simulation}. The current seller $\mathcal{S}_{crnt}$ and consumers $\mathcal{C}_{crnt}$ list, along with the updated prices, are provided to {\rm EM} algorithm. Based on the output, the final matching list $\mu_{final}$ and price list $price$ are updated. After $T$ iterations, {\rm NEM} algorithm terminates with $\mu_{final}$ containing pairs of sellers and consumers who will trade energy, and $price$ containing price at which energy will be traded. Those sellers (or consumers) who have not been matched can sell (or buy) energy from the electricity distribution company. \vspace{-2mm}
\begin{prop}[\textbf{Computational Complexity}] The {\rm NEM} algorithm converges with computational complexity of $\mathcal{O}(TNML_{\rm max}{\tau}_{\rm max})$, where $T$ is number of iterations in {\rm NEM}. $N$ and $M$ are the number of consumers and sellers, respectively. $L_{\rm max}$ is the maximum number of blocks demanded by any consumer, and ${\tau}_{\rm max}$ is the maximum time taken by any seller to allocate blocks to a consumer in the network.
\label{prop:3}
\end{prop} \vspace{-2mm}
\noindent Since, the {\rm NEM} algorithm repeatedly runs {\rm EM} algorithm for $T$ iterations, so computation complexity of {\rm NEM} algorithm is $T$ times the complexity of {\rm EM} algorithm.
 \vspace{-2mm}
\begin{prop}[\textbf{Feasibility and Stability}] The matching result of the {\rm NEM} algorithm is both feasible and stable.
\end{prop} \vspace{-2mm}
\noindent\textbf{Feasibility.} The {\rm NEM} algorithm uses {\rm EM} algorithm for matching sellers with consumers. Since output matching of {\rm EM} algorithm is feasible (refer proposition \ref{prop:2}), consequently matching from {\rm NEM}  algorithm is also feasible.

\noindent\textbf{Stability.} As in Proposition \ref{prop:2}, the matching of {\rm EM} algorithm is stable, so matching achieved by {\rm NEM} algorithm at the end of an iteration is also stable. However, in the next iteration of {\rm NEM} algorithm, sellers and consumers approach each other with modified selling and buying prices. This can interfere with the earlier matching of sellers and consumers, who have agreed to trade energy at old prices. To remove this issue, in {\rm NEM} algorithm, sellers and consumers who have agreed to trade in a given iteration must trade energy before the start of the next iteration. Before the start of the next iteration of {\rm NEM} algorithm,  sellers and consumers lists are updated by removing the above-discussed seller and consumers. Hence, matching obtained using {\rm NEM} algorithm is also stable. 

\noindent\textbf{Prices offered by sellers and consumers.} At the end of each iteration, the sellers and consumers negotiate energy prices by changing their selling and buying prices. Initial prices offered by sellers and consumers depend on their monetary and energy requirements, respectively. Sellers and consumers with immediate requirements will offer low selling and high buying prices, respectively, in the first round. Due to the prices offered, the chances are they will get matched in the first iteration. Those sellers and consumers who have moderate requirements will offer moderate prices. The chances are that they will get matched in the middle iterations. Those sellers with sufficient storage and no immediate monetary needs will offer energy at high selling prices. Similarly, consumers who can store energy with no immediate energy requirements will offer low buying prices in the first iteration. Chances are such sellers and consumers will be matched in the last iteration or may remain unmatched.\\
\textbf{{NEM} and {EM} application.} Although {\rm NEM} reduces energy prices for consumers, but sellers will also be receiving lesser energy prices. Thus consumers will prefer {\rm NEM}, while sellers will prefer {\rm EM} algorithm. Apart from price difference, the execution time of {\rm NEM} is higher than {\rm EM} algorithm\footnote{In a single iteration of {\rm NEM} algorithm, {\rm EM} algorithm gets executed multiple time. }. So a consumer who has urgent energy requirements will prefer {\rm EM} algorithm. Thus some users may prefer {\rm EM}, and some may prefer {\rm NEM} algorithm. Hence some P2P markets may employ {\rm EM}, and others may employ {\rm NEM} algorithm for grouping. This situation is analogous to the co-existence of fixed-price shops, with shops allowing price negotiation in society.
 \vspace{-3mm}
\section{Performance Analysis}
 \vspace{-1mm}
\label{sec:performance}
This section first discusses the simulation setup, negotiation process, and then how the final price of energy is calculated. After that, we simulate $4$ algorithms that group sellers with consumers. First, the proposed algorithms, \textit{viz.}, energy matching {\rm EM} and negotiable energy matching {\rm NEM} are simulated to determine the effect of negotiation on energy prices. Then the performance of these algorithms is compared with existing ones: double auction (\rm DAM) \cite{Energy_Coin} and first come first serve (\rm FFS) \cite{dec_smartgrid}. We determine 1. loss in sellers' revenue due to central authority acting as an intermediary and 2. change in selling and buying prices due to malicious users. We now describe the simulation network.
 \vspace{-3mm}
\subsection{Simulation network}
 \vspace{-1mm}
\label{subsec:simulation} We analyze a {\rm P2P} network with $45$ sellers and $45$ consumers. Since this paper focuses on group formation, the calculation of bids from sellers' and consumers' is not discussed. We assume that sellers are equally divided into five groups, with each group having a different bid than the other. Sellers bids for a single block is given by \vspace{-2mm}
 \begin{equation}
\label{eq:seller_bid}
 \vspace{-2mm}
\mathbf{\bar{b}_{\rm sel}}=\{b_s-2\Delta_s,b_s-\Delta_s,b_s,b_s+\Delta_s,b_s+2\Delta_s\},
\end{equation} where $i^{th}$ element of vector is equal to bid of user in the $i^{th}$ group. $b_s$ is the average bid price quoted by all the sellers, and $\Delta_s$ is the difference in bid price of adjacent groups of sellers. A block is ${\rm 1KWh}$ of energy.  Similarly, consumers are equally divided into $5$ groups with bids for a different groups as   \vspace{-2mm}
\begin{equation}
	\label{eq:consumer_bid}
	 \vspace{-2mm}
\mathbf{\bar{b}_{\rm con}}=\{b_c-2\Delta_c,b_c-\Delta_c,b_c,b_c+\Delta_c,b_c+2\Delta_c\}.
\end{equation}
 $b_c$ is the average bid price quoted, and $\Delta_c$ is the difference between adjacent bids of the groups. Consumers approach sellers with their respective bids, and groups are formed between them using {\rm EM} or {\rm NEM} algorithm. {\rm NEM} algorithm allows price negotiation by adjusting bids after each iteration. The bids are adjusted as follows. \\
 \textbf{Adjustments of bids during negotiation} At end of each iteration a seller $s_j$ decrease its bid by $\delta_{s_j}$ and consumer $c_i$ increase its bid by $\delta_{c_i}$. The sellers and consumers can modify their bids, till it reaches minimum or maximum value, respectively. If total number of iterations allowed is $T$, then  $\delta_{s_j}$ and $\delta_{c_i}$ are calculated as follows.\\
 \textbf{$\boldsymbol{\delta_{s_j}}$ calculation.} A seller can sell energy locally or to electricity distribution companies. The energy storage devices in the electricity distribution company are generally located far from the seller. Losses are associated with energy transmission. Such losses are minimum when energy is traded locally. Hence, local consumers will receive more energy than electricity distribution companies for the same amount of energy sold by a seller. So the buying price offered by local consumers will be more than the electricity distribution company. Hence, the minimum selling price equals the price an electricity distribution company offers to purchase energy. We define $\delta_{s_j}$ in such a way that the selling price of a seller $s_j$ decreases equally in each iteration, and is equal to the minimum selling price in the last iteration. Thus,  
 \begin{equation}
 	\label{eq:selling_incriment} 
 	\delta_{s_j}=\frac{{\rm Initial~Selling~Price~of~} s_j-{\rm Minimum~Selling~Price}}{T-1}.
 \end{equation}
 \noindent\textbf{$\boldsymbol{\delta_{c_i}}$ calculation.} Consumers can purchase energy locally or from an electricity distribution company. Due to transmission losses, energy will be available at lower prices in the local network. Therefore, Maximum Buying Price = Selling price an electricity distribution company offers. For simulation, we set $\delta_{c_i}$ such that the buying price of a consumer $c_i$ increases equally in each iteration, so that in the last iteration, its buying price is equal to the maximum buying price. Thus,  
 \begin{equation}
 	\label{eq:buying_incriment}
 	\delta_{c_i}=\frac{{\rm Maximum~Buying~Price}-{\rm Initial~Buying~Price~of~} c_i}{T-1}
 \end{equation}
 During simulation, number of iterations $T$ is set to $6$. For $6$ iterations, a good trade-off between running time of an algorithm and negotiated prices is achieved. Once the bids are finalized, the trading starts. The trading price of energy during in {\rm EM} and {\rm NEM} algorithm is calculated as follows.\\
 \textbf{Trading price of energy}
\label{sec:negotiation}In {\rm EM} algorithm, if the consumer's bid is lesser than the seller's bid, then energy is traded at the seller's bid. Else, the energy is traded at the average bid price proposed by the consumer and seller.
\\  In {\rm NEM} algorithm, if the iteration is not the last, then the energy is traded only when the consumer's bid is higher than the seller's bid. 
The trading price is the average bid price proposed by the seller and the consumer. In the last iteration, the trading price is determined in the same way as {\rm EM} algorithm. 
 \vspace{-3mm}
\subsection{Effect of negotiation on trading price of energy}
 \vspace{-1mm}
%
%
We compare the prices at which energy is traded in {\rm EM} and {\rm NEM} algorithm. Let seller's supply and the consumer's block demand can vary between $1$ to $5$ blocks. The supply and demand, are randomly generated with a uniform probability distribution. For the blocks available with seller, the consumers will submit their bid to fulfill their requirement. For ease of analysis, let us assume following bid parameters in equations \eqref{eq:seller_bid} and \eqref{eq:consumer_bid}. Average bid price of sellers $b_s$ and consumers $b_c$  are equal to $0.8$, whereas difference in bids of adjacent groups of sellers $\Delta_s$ and  consumers $\Delta_c$ are equal to $0.1$.

In {\rm EM} algorithm, bids of sellers and consumers are fixed, whereas {\rm NEM} algorithm they can vary. Consumers increase their bids in each iteration,and it reaches its maximum value in last iteration; whereas sellers decrease their bids in subsequent iterations till it reaches a minimum value. On changing these maximum and minimum limits, the price at which energy is traded changes, see Fig. \ref{fig:EM_NEM}. We have simulated {\rm NEM} algorithm for $5$ different set of limits (maximum bid by the consumer, and minimum bid by the seller), \textit{viz.}, $10,1$, $10,6$, $20,1$, and $20,6$, and compared it with {\rm EM} algorithm. Due to negotiation, the price at which energy is traded is lesser in {\rm NEM} than {\rm EM} algorithm (Fig. \ref{fig:EM_NEM}). In {\rm NEM} algorithm, if the maximum limit on a bid by the consumer is increased, the trading price of energy increases, and if the limit on the minimum bid by a seller decreases, then the trading price decreases. Trading price is the average of bids by a seller and a consumer. Increasing the limit on a maximum bid by a consumer makes it possible for them to bid more in subsequent iterations, thereby increasing the trading price. Decreasing the limit on a minimum bid by a seller makes it possible for them to bid less in subsequent iterations, thereby decreasing trading prices. \vspace{-3mm}
\subsection{Comparison of the proposed and existing algorithms}
\pgfplotstableread[row sep=\\,col sep=&]{
	interval & trade  \\
	EM   & 0.85 \\
	$\text{NEM}_{10,1}$& 0.76 \\
	$\text{NEM}_{10,6}$ &0.79 \\
	$\text{NEM}_{20,1}$ &0.82 \\
	$\text{NEM}_{20,6}$ & 0.84 \\
}\comparison
\pgfplotstableread[row sep=\\,col sep=&]{
	interval & sell & buy \\
	DAM    & 0.03 & 0.23 \\
	FFS     & 0.05 & 0.38 \\
	EM    &0.14  &0.14  \\
	NEM   &0.13 &0.13  \\
}\incentive
%
\begin{figure*}[ht!]
	\hspace{-3.cm}
	\centering 	 
	\begin{subfigure}[h]{0.35\textwidth}
		\centering		 	
	\vspace*{-3mm}\begin{tikzpicture}[scale=0.60]
		\begin{axis}[
			ybar,
			bar width=0.4cm,
			symbolic x coords={EM, $\text{NEM}_{10,1}$, $\text{NEM}_{10,6}$, $\text{NEM}_{20,1}$, $\text{NEM}_{20,6}$},
			xtick=data,
			nodes near coords,
			nodes near coords align={vertical},
			ymin=0,ymax=1,
			ylabel style={text width=6cm},
			ylabel={\textbf{\hspace{6.5mm}Average price$~\boldsymbol{\left(\frac{\rm dollar}{\rm KWh}\right)}$}},
			xlabel={\textbf{Matching algorithm}}
			]
			\addplot [black,fill=blue]table[x=interval,y=trade]{\comparison};
			
		\end{axis}
	\end{tikzpicture}
	\captionsetup{justification=centering}\vspace*{-2mm}	\caption{Energy  price using {\rm EM} and different instances of {\rm NEM} algorithm.}
		\textsuperscript{\hspace{-1mm}$\text{NEM}_{x,y}$: $x$ is maximum bid by a consumer,}\\\textsuperscript{ and $y$ is minimum price demanded by a seller.}
	\label{fig:EM_NEM}
	\end{subfigure}
	\begin{subfigure}[h]{0.35\textwidth}
		\centering		 	
	\vspace*{-12.9mm}\begin{tikzpicture}[scale=0.60]
		\begin{axis}[
			ybar,
			bar width=0.4cm,
			symbolic x coords={DAM,FFS,EM,NEM},
			xtick=data,
			ymin=0,ymax=1,
			ylabel style={text width=6cm},
			ylabel={\textbf{\hspace{6.5mm}Average price$~\boldsymbol{\left(\frac{\rm dollar}{\rm KWh}\right)}$}},
			xlabel={\textbf{Matching algorithm}}
			]
			\addplot [black,fill=red]table[x=interval,y=sell]{\incentive};
			\addplot [black,fill=yellow]table[x=interval,y=buy]{\incentive};
			\legend{\textbf{Selling Price}, \textbf{Buying Price}}
		\end{axis}
	\end{tikzpicture}
	\vspace*{-2mm}
	\caption{ Seller's revenue loss = Buying - Selling price.}
	\label{fig:incentive}
	\end{subfigure}
	\begin{subfigure}[h]{0.3\textwidth}
		\centering		 	
			\vspace*{-9mm}\begin{threeparttable}
			\resizebox{7cm}{!}{
				\begin{tabular}[b]{ C{2cm}C{1.5cm}C{1.5cm}C{1.5cm}C{1.5cm}} 
					\hline
					\hline
					Central authority's behavior & \multicolumn{4}{c}{
						Price: selling, buying ${\left(\frac{\rm dollar}{\rm KWh}\right)}$} \\
					& DAM  &FFS & EM  & NEM  \\
					\hline
					No maliciousness &0.033, 0.232&0.054, 0.382&0.136, 0.136&0.131, 0.131 \\
					\rowcolor{Gray} 
					Seller preference&0.059, 3.000&0.062, 0.443&0.136, 0.136&0.131, 0.131\\
					Consumer preference&0.031, 0.200&0.001, 0.007 &0.136, 0.136 &0.131, 0.131\\
					\hline
					\hline
				\end{tabular}	}
			\captionsetup{justification=centering}
			\vspace*{-4mm}
			\caption{ Price variation when central authority act maliciously.}
			\end{threeparttable}
			\centering
				\label{fig:sim_dynamic}
			\end{subfigure}
		\vspace*{-5mm}
	\caption{Comparison of different matching algorithms}\vspace*{-5mm}
	\label{fig_comparison}
\end{figure*}
The performance of the {\rm EM} and {\rm NEM} algorithms is compared with the following algorithms.\\ 
\textbf{Double Auction Mechanism {(DAM)} \cite{Energy_Coin}.} {\rm DAM} uses the auctioneer to group sellers with consumers. An auctioneer is a user in the network with high computational power. Let there be $I$ consumers and $J$ sellers in the network. To participate in trading, they have to register with the auctioneer $n$. Consumers and sellers submit their initial bids to the auctioneer. The auctioneer $n$ consolidates bids of consumers and sellers in matrices $\mathbf{B^{n}}=[b_{ij}^n]$ and $\mathbf{S^{n}}=[s_{ij}^n]$, of sizes $I{\times}J$ and $J{\times}I$ respectively. $b_{ij}^n$ is the price offered (bid) by consumer $i$ to seller $j$, and $s_{ij}^n$ is the price demanded (bid) by seller $i$ from consumer $j$.

Based on the initial bids, the auctioneer generates the demand and supply profile for consumers and sellers: How much energy any consumer $i$ will demand from different sellers and how much energy any seller $j$ will offer to different consumers. These profiles are generated in such a way to maximize social welfare (\textit{i.e.}, sum of utilities of participants in trading). To maximize social welfare, auctioneer solves the following optimization problem \cite{Energy_Coin}. 
\begin{equation}
	\label{eq:optim} 
	\max_{\mathbf{{C}^{n}},\mathbf{{D}^{n}}}\sum_{i=1}^{I}\sum_{j=1}^{J}\left[b_{ij}^{n}{\rm ln}c_{ij}^{n}-s_{ji}^{n}d_{ji}^{n}\right],
\end{equation}
where $\mathbf{{C}^{n}}=[c_{ij}^n]$ and $\mathbf{{D}^{n}}=[d_{ij}^n]$ are demand and supply matrices of sizes $I{\times}J$ and $J{\times}I$, respectively. $c_{ij}^n$ is the energy demanded by $i^{\rm th}$ consumer from $j^{\rm th}$ seller, and $d_{ij}^n$ is the energy offered by $i^{\rm th}$ seller to $j^{\rm th}$ consumer. The sum of $b_{ij}^{n}{\rm ln}c_{ij}^{n}$ denotes the satisfaction derived by all the users on charging and sum of $s_{ji}^{n}d_{ji}^{n}$ denotes the overall cost incurred in energy generation. To maximize social welfare, user satisfaction should be maximum, and energy generation cost should be minimum.

 After solving \eqref{eq:optim}, auctioneer sends demand $\mathbf{{C}^{n}}$ and supply $\mathbf{{D}^{n}}$ matrices to consumers and sellers, respectively. Based on demand matrix $\mathbf{{C}^{n}}$, consumer $i$ update its bid to seller $j$ as 
\begin{equation}
	\label{eq:buying_price} 
	b_{ij}^n=\frac{{\eta}{\zeta }c_{ij}^n}{\left(\eta\sum_{j=1}^{J}c_{ij}^n-c_{i}^{n,{\rm min}}+1\right){STO}_i^n},
\end{equation}  
where $\eta=0.8$ is the average charging efficiency, and ${STO}_i^n=0.2$ is the energy state of consumer $i$ before charging. $\zeta=0.6$ is a constant, and $\frac{\zeta}{{STO}_i^n}$  denotes the willingness of a user to charge.  $c_{i}^{n,{\rm min}}=5$ is the minimum energy demanded by a consumer.
Similarly, based on supply matrix $\mathbf{{D}^{n}}$, seller $j$ update its price bid for energy as \vspace{-2mm}
\begin{equation}
		\label{eq:selling_price} \vspace{-2mm}
	s_{ji}^n=2l_1d_{ji}^n+l_2.
\end{equation} 
where $l_1$ and $l_2$ are constants set to 0.01 and 0.015, respectively. 

After calculating their bids, sellers and consumers send these to the auctioneer. The auctioneer, consolidates these bids to generate new bid matrices $\mathbf{B^{n}}$ and $\mathbf{S^{n}}$, for consumers and sellers, respectively. Let $\mathbf{B^{n(t)}}$ and $\mathbf{S^{n(t)}}$ be bid matrices generated by auctioneer at time $t$, and $\mathbf{B^{n(t-1)}}$ and $\mathbf{S^{n(t-1)}}$ are generated in previous time $t-1$. Let $RDB=\frac{\left|\sum b_{ij}^{n(t)}-\sum b_{ij}^{n(t-1)}\right|}{\sum b_{ij}^{n(t)}}$ and $RDS=\frac{\left|\sum s_{ji}^{n(t)}-\sum s_{ji}^{n(t-1)}\right|}{\sum s_{ji}^{n(t)}}$. If $RDB$ and $RDS$ are less than $0.001$, then new and  old bids converge, and algorithm terminates. Otherwise, new bids $\mathbf{B^{n}}$ and $\mathbf{S^{n}}$ are used in \eqref{eq:optim} to generate new supply $\mathbf{C^{n}}$ and demand $\mathbf{D^{n}}$ matrices. These matrices are sent to consumers and sellers, respectively, and same process is repeated until, convergence criterion is met.\\
\textbf{First come first service (FFS) \cite{dec_smartgrid}.} In first come first service, sellers post their bid price and energy available on the auctioning board hosted by an electricity distribution company. After reading bids, a consumer approach seller whose price bid matches with its bid price. Seller trades energy with the consumer whose bid is received first.
 \vspace{-1mm}
\subsubsection{Normalization of parameters in algorithms for comparison} In {\rm DAM} algorithm, the auctioneer fixes the supply and demand of sellers and consumers, respectively, in \eqref{eq:optim}. Based on supply and demand, bids of sellers and consumers are calculated in \eqref{eq:buying_price} and \eqref{eq:selling_price}. For comparison of algorithms, all the parameters except bids from the {\rm DAM} algorithm are used. 

As bids in {\rm DAM} are controlled, sometimes bids of sellers and consumers may be very close to each other. To simulate a real-market scenario, we have created a difference in the bids while simulating the remaining algorithms {\rm FFS, EM, and NEM}. However, the average bids of sellers and consumers remain the same during all algorithms' simulations. For the remaining algorithms, the bid parameters in equation \eqref{eq:seller_bid} and \eqref{eq:consumer_bid} are as follows. The average bid prices for sellers $b_s$ and consumer $b_c$ are equal to $avg_{\rm sel}$ and $avg_{\rm con}$, respectively.  $avg_{\rm sel}$ and $avg_{\rm con}$ are average bid prices of sellers and consumers, respectively, from simulation of {\rm DAM} algorithm. Difference between bids of adjacent groups for sellers $\Delta_s$ and consumers $ {\Delta_c}$ are equal to $\frac{avg_{\rm sel}}{10}$ and $\frac{avg_{\rm con}}{10}$, respectively. 
\subsubsection{Revenue loss due to intermediary} 
\pgfplotstableread[row sep=\\,col sep=&]{
	interval & sell & buy \\
	DAM    & 0.03 & 0.23 \\
	FFS     & 0.05 & 0.38 \\
	EM    &0.14  &0.14  \\
	NEM   &0.13 &0.13  \\
}\incentive
%
The central authority (auctioneer in {\rm DAM} and electricity distribution company in {\rm FFS}) facilitate group formation and charge for their service. In {\rm DAM} algorithm,  the payment (buying price) made by a consumer $i$ is given by 
\begin{equation}
	Pay_i(\mathbf{B_i^n})=\sum_{j=1}^{J}b_{ij}^n,
\end{equation}
where $J$ is the total number of sellers with auctioneer $n$.A portion of the payment by the consumer is transferred to the auctioneer. The reward (selling price) of the seller $j$ is different from the consumer payment and is given by 
\begin{equation}
	Rwd_j(\mathbf{S_j^n})=\sum_{i=1}^{I}\frac{(s_{ji}^n)^2}{4l_1}+r^{\rm min}_j,
\end{equation}
where $I$ is the total number of consumers registered with the auctioneer. $l_1=0.01$ is constant, and $r^{\rm min}_j=0.01$ is initial reward given by auctioneer $n$ to seller $j$ for registering with $n$. For normalization of incentive, the ratio between average selling and buying price is kept same in both {\rm DAM} and {\rm FFS}. Since central authority is charging for its service, there is a loss of revenue for sellers in {\rm DAM} and {\rm FFS}. The money paid by consumers is not entirely getting transferred to sellers (Fig. \ref{fig:incentive}). Whereas for decentralized algorithms {\rm EM} and {\rm NEM}, there is no loss of revenue.  
\subsubsection{Impact due to  Malicious behavior of central authority}
%
%
In {\rm DAM} algorithm, the supply and demand of sellers and consumers, respectively, are set by a central authority (auctioneer). The auctioneer can behave maliciously and manipulate supply and demand to benefit certain sellers and consumers. To demonstrate how certain sellers can profit, let us consider that the auctioneer sets supply from $15$ preferred sellers to a maximum value $20 ~{\rm KWh}$ and supply from remaining sellers is $0$. The bids are dependent on supply and demand, refer \eqref{eq:buying_price},\eqref{eq:selling_price}. Since supply is less (only $15$ sellers are involved in trading), the average price at which energy is traded increases (Fig. 2c), thereby profiting the sellers promoted by the auctioneer. To demonstrate how certain consumers can profit, let us consider that the auctioneer sets demand of $15$ preferred consumers to a maximum value $20 ~{\rm KWh}$ and remaining consumers $0$. Since demand is less, the average energy prices fall, profiting selected consumers (Fig. 2c). In {\rm FFS}, a central authority can collude with sellers to give them an advantage. We consider a scenario where the central authority promotes $15$ most expensive sellers. It hides the bids of remaining sellers on the auction board. This forces consumers to buy from the most expensive sellers. Consequently, both selling and buying prices increase (Fig. 2c). The central authority can also collude with consumers. We simulate a scenario where the central authority favours $15$ consumers. It tries to match these consumers with sellers that quote minimum price bids by not sharing trading data with other consumers. Consequently, these buyers can purchase energy at lower prices (Fig. 2c). In {\rm EM} and {\rm NEM} algorithms, there is decentralized group formation, so their performance is immune to malicious behavior.
\subsubsection{Comparison of execution time} 
In the worse case of a {\rm FIFO} algorithm, a single user will contact all the sellers once for energy. If there are $N$ consumers and $M$ sellers, then the computational/time  complexity  of the algorithm is $\mathcal{O}(NM)$. The time complexities of {\rm EM} and {\rm NEM} algorithms are $\mathcal{O}(NML_{\rm max}{\tau}_{\rm max})$ and $\mathcal{O}(TNML_{\rm max}{\tau}_{\rm max})$, respectively (refer propositions \ref{prop:1} and \ref{prop:3}). $L_{\rm max}$, ${\tau}_{\rm max}$, and $T$ are the maximum number of blocks demanded, maximum block allocation time, and total iterations in {\rm NEM} algorithm, respectively. Clearly, {\rm FIFO}'s time complexity is lesser than the proposed algorithms. 
	
{\rm DAM} algorithm may take less or more time than the other algorithms. It continues to execute until the current round bids converge to previous round bids. The number of rounds required for convergence is not fixed and changes with variations in the number of sellers and consumers and their initial bids. For the simulation setting used in this paper, we observed that the run time of {\rm DAM} was higher than the proposed algorithms. The run time of algorithms {\rm FFS}, {\rm EM}, {\rm NEM}, and {\rm FFS} are $1.2$, $8.76$, $60.11$, and $317.14$ seconds, respectively. These results are consistent with time complexities of algorithm calculated earlier.

\section{Conclusion}
\label{sec:conclusion}
We have proposed two distributed algorithms, \textit{viz.,} energy matching {(\rm EM)} and negotiable energy matching ({\rm NEM}) for matching sellers with consumers in {\rm P2P} energy trading. Sellers and consumers cannot negotiate prices in {\rm EM} algorithm. So we have extended it and presented {\rm NEM} algorithm that allows price negotiation. It was also proved that these algorithms converge in real-time and produce a stable matching. On comparing them with existing ones, results demonstrated no influence of centralized control on the prices in proposed algorithms. Further, energy was traded at lower prices in the network employing {\rm EM} or {\rm NEM} algorithm when compared with the existing ones. Among {\rm EM} and {\rm NEM} algorithm, the latter provides lower energy prices due to price negotiation, but it requires a higher time for execution. 

The proposed work can establish a free market that can provide energy at cheaper rates. But this work does not provide any mechanism that can decide initial bids of sellers and consumers. This work can be extended to include the economic analysis that uses demand and supply of energy to determine initial bids.
\bibliographystyle{IEEEtran}
\bibliography{ref}

\ifCLASSOPTIONcaptionsoff
  \newpage
\fi

\end{document}